\documentclass[prl,twocolumn,showpacs,preprintnumbers,amsmath,amssymb]{revtex4}
\usepackage{graphicx}
\usepackage{dcolumn}
\usepackage{bm}
\usepackage{psfig}
\newcommand {\be}{\begin{eqnarray}}
\newcommand {\ee}{\end{eqnarray}}

\begin{document}
\title{Multiple condensed phases in attractively interacting Bose systems}
\author{M. M{\"a}nnel$^{1}$, K. Morawetz$^{2,3}$, P. Lipavsk\'y$^{4,5}$
}
\affiliation{$^1$Institute of Physics, Chemnitz University of Technology, 09107 Chemnitz, Germany}
\affiliation{$^2$International Center of Condensed Matter Physics, University of Brasilia, 70904-970, Brasilia-DF, Brazil}
\affiliation{$^3$Research-Center Dresden-Rossendorf, Bautzner Landstr. 128, 01328 Dresden, Germany}
\affiliation{$^4$Institute of Physics, Academy of Sciences, Cukrovarnick\'a 10, 16253 Prague 6, Czech Republic}
\affiliation{$^5$Faculty of Mathematics and Physics, Charles University, Ke Karlovu 3, 12116 Prague 2, Czech Republic}

\begin{abstract}

We investigate a Bose gas with finite-range interaction using a scheme to eliminate unphysical processes in the T-matrix approximation. In this way the corrected T-matrix becomes suitable to calculate properties below the critical temperature.  For attractive interaction, an Evans-Rashid transition occurs between a quasi-ideal Bose gas and a BCS-like phase with a gaped dispersion. The gap decreases with increasing density and vanishes at a critical density where the single-particle dispersion becomes linear for small momenta indicating Bose-Einstein condensation. The investigation of the pressure shows however, that the mentioned quantum phase transitions might be inaccessible due to a preceding first order transition.
\end{abstract}

\date{\today}
\pacs{
03.75.Hh, 
03.75.Nt,
05.30.-d, 
64.10.+h 
}
\maketitle

The multiple-scattering correlations collected in terms of the T-matrix are a powerful tool to describe properties of interacting many-body systems. Compared to the T-matrix of the scattering theory, the many-body T-matrix contains pairs of medium-dependent intermediate propagators. They describe the influence of the surrounding medium and their selfconsistent determination is a challenging task. It was first studied in dense Fermi systems, where taking the particle-particle correlations as well as hole-hole correlations into account a combination of self-consistent propagators appear. The application of this fully self-consistent approach called the Galitskii-Feynman approximation below the critical temperature of condensation is questionable. Considering the case of cooper pairing the break down of this approximation can be observed by the fact that although the divergence of the T-matrix leads to the correct critical temperature the Galitskii-Feynman approximation does not lead to the correct gap dispersion for the condensed phase \cite{L08}.

It was observed by Kadanoff and Martin \cite{KM61} and used later on \cite{MBL99,HCCL07,L08,SLMM09} that an asymmetric breaking of the selfconsistency in the T-matrix, such that one propagator is used selfconsistently and the other non-selfconsistently, leads immediately to the correct gap dispersion while the symmetric selfconsistent one does not. This so called Prange paradox \cite{Prange60} remains puzzling since a seemingly worse approximation leads to better results. Recently it was found that the reason for that could be unphysical repeated collisions \cite{L08}.

While in the thermodynamic limit the contributions of these or similar unphysical processes usually vanish, since each channel contributes separately with the weight of inverse volume, they become essential when a condensed phase appears since then the occupation of one particular channel is proportional to the volume itself. The advantage of eliminating only the contributions of single channels as proposed in Refs.~\cite{L08} and \cite{SLMM09} is that the formation of pairs and their condensation can be described within the same approximation. We note that the proposed elimination shemes yield similar results as the often used mean field approximations with anomalous functions \cite{SG98}. Further improvements of the T-matrix can be achieved if the interference of particle-particle channels and particle-hole channels are treated as noted e.g. in \cite{SK06}.

From these experiences which gives us the possibility to treat the condensed and non-condensed phase on the same footing we expect to gain more insight also to the problem of interacting Bose systems. As a prominent example, currently much debated, we present the application to these systems and show the appearance of `inaccessible' condensed phases. 
The theory predicts that a Bose gas with contact interaction and a scattering length $a_0<0$, i.e., attractive interaction,
undergoes the so called Evans-Rashid transition \cite{ER73,S94}, which is the boson analogue to the BCS transition for fermions. It has been shown by Stoof \cite{S94}, however, that such Bose gas becomes liquid or solid \cite{FW71,S94} before the Evans-Rashid transition or Bose condensation can appear. On the other hand a BCS-like molecular Bose-condensation as well as a condensation of both atoms and molecules is especially predicted near a Feshbach resonance \cite{RPW04, RDSS04,RWP08,KMDS09}. Here we find three quantum phases: (I) An ideal gas with or without bound states, (II) a gaped phase and (III) the Bose-Einstein condensate. Whether the latter two phases are experimentally accessible is however unclear so fare.

Let us assume a homogeneous Bose gas with an attractive interaction strong enough to form a two-particle bound state. Such bound states can be found as poles of the T-matrix, which reads in ladder approximation
\be
{\cal T}_{q}\!(iz,\!p,\!k)\!=\!V_{q}\!(p,\!k)\!-\!\frac{1}{\Omega}\!\sum_{k'}V_{q}(p,\!k') {\cal G}_q(iz,\!k'){\cal T}_{q}(iz,\!k',\!k),
\label{tmat}
\ee
i.e., the Bethe-Salpeter equation, where $q$ is the center of mass momentum and 
\be
{\cal G}_q(iz,p)=T\sum_{\omega}G\left(i\omega,p\right)G\left(iz-i\omega,q-p\right)
\ee
is the two-particle propagator. The Matsubara frequencies $\omega$ are even multiples of $\pi T$ and $\Omega$ is the system volume. By closing the loop, including exchange and summing over all $q$ we obtain the selfenergy
\be
\Sigma(i\omega,k)&=&-\frac{T}{\Omega}\sum_{z,q}\left[{\cal T}_{q}(iz,k,k)+{\cal T}_{q}(iz,k,q-k)\right]\nonumber\\
&&\times G(iz-i\omega,q-k).
\label{ise}
\ee
For fermions this is called the Galitskii-Feynman approximation. We use the sign convention of Ref.~\cite{AGD75}. The set of equations is closed by the Dyson equation 
\be
G=G^0+G^0\Sigma G
\label{Dyson}
\ee 
with the free Green function, ${G^0=1/[i\omega-\epsilon_0(k)]}$ and the free single-particle energy $\epsilon_0(k)={\hbar^2k^2}/{2m_0}-\mu_0$.

Let us assume a bound state with momentum $q$ and energy $\omega_q$, i.e., {$1/{\cal T}_{q}(\omega_q-2\mu,p,k)=0$}. The chemical potential $\mu$ will be specified later. We further assume that the system has excitations at energies $E_k^\pm$, represented by poles of the dressed Green function, i.e., $1/G(E_k^\pm,k)=0$. With the Feynman trick we convert the sum over Matsubara frequencies in Eq.~(\ref{ise}) into a sum over the poles of the T-matrix and the dressed Green function \cite{M90} which leads to
\be
\Sigma(i\omega,k)&=&\sum_q\Delta_q^2(k)G(\omega_q-2\mu-i\omega,q-k)\nonumber\\
&&+\sum_{q,\pm} n_{q-k}^\pm\left[{\cal T}_{q}(E_{q-k}^\pm+i\omega,k,k)
\right.\nonumber\\&&\left.
+{\cal T}_{q}(E_{q-k}^\pm+i\omega,k,q-k)\right],
\ee
where
\be
&&\Delta_q^2(k)=\frac{f_B(\omega_q-2\mu)}{\Omega}\nonumber\\
&&\times{\rm Res}\left[{\cal T}_{q}(z,k,k)+{\cal T}_{q}(z,k,q-k),\omega_q-2\mu\right]
\label{dq2}
\ee
and
\be
n_k^\pm=\frac{f_B(E_k^\pm)}{\Omega}{\rm Res}[G(z,k),E_k^\pm].
\ee
The Bose distribution is ${f_B(\epsilon)=[\exp(\epsilon/T)-1]^{-1}}$, the residuum of function $h(z)$ at its pole $\epsilon$ is denoted by ${\rm Res}[h(z),\epsilon]$. 
The chemical potential $\mu$ increases with density until it approaches the lowest bounded state, which is the state of zero momentum, $q=0$. At this point, $2\mu=\omega_0$, the Bose distribution becomes divergent with a value proportional to the system volume, i.e., the Evans-Rashid transition which means the onset of a condensation of zero momentum bound states. The divergent Bose distribution results in a singular contribution of the $(q,\omega_q-2\mu)=(0,0)$ channel to the selfenergy
\be
\Sigma_{\rm sin}(i\omega,k)=\Delta_0^2(k)G(-i\omega,-k).
\label{s0gf}
\ee
If the interaction allows for several bound states the singular contribution will always come from the one with lowest energy. All the other contributions to the selfenergy are regular being proportional to $1/\Omega$.
We use a quasi-particle approximation ${\Sigma_{\rm reg}(i\omega,k)=\Sigma(i\omega,k)-\Sigma_{\rm sin}(i\omega,k)\approx
\Sigma_{\rm reg}[\epsilon(k),k]}$ with the quasi-particle energy
\be
\epsilon(k)=
\epsilon_0(k)+\Sigma_{\rm reg}[\epsilon(k),k]=\frac{\hbar^2k^2}{2m}-\mu+{\cal O}(k^3).
\label{qp}
\ee
Expansion for small momenta determines an effective chemical potential $\mu$ and an effective mass $m$. The quasiparticle damping could be considered as well if the imaginary part of the self energy is taken into account \cite{SHK08}. For the present message we neglect this effect.

As argued in Refs.~\cite{L08} and \cite{SLMM09}, the Galitskii-Feynman approximation is not able to give the correct dispersion in the condensed phase. To avoid this problem the loop of the singular self energy contribution (\ref{s0gf}) has to be constructed with reduced Green functions which do not include the singular self energy
\be
G_{\not\Delta}=G^0+G^0\Sigma_{\rm reg}G_{\not\Delta}\approx\frac{1}{i\omega-\epsilon(k)},
\label{gr}
\ee
i.e.,
\be
\Sigma_{\rm sin}(i\omega,k)=\Delta_0^2(k)G_{\not\Delta}(-i\omega,-k)
\label{s0}
\ee
and also the two particle propagator of the $(0,0)$ channel has to be constructed with one reduced line
\be
{\cal G}_0(0,p)=T\sum_{\omega}G\left(i\omega,p\right)G_{\not\Delta}\left(-i\omega,-p\right).
\label{g2pns}
\ee
The dressed Green function follows from the Dyson equation (\ref{Dyson}), Eqs.~(\ref{gr}) and (\ref{s0}) as
\be
G=G_{\not\Delta}+G\Sigma_{\rm sin}G_{\not\Delta}\approx\frac{i\omega+\epsilon(k)}{(i\omega)^2-E^2_k},
\label{gdsc}
\ee
with a two-branches quasiparticle dispersion $E_k^\pm=\pm E_k$,
\be
E_k=\sqrt{\epsilon^2(k)-\Delta_0^2(k)}.
\label{dispersion}
\ee
This presents a closed set of equations again. With the Green function (\ref{gdsc}) we can now calculate 
the total density
\be
n=\!-\frac{T}{\Omega}\!\sum_{\omega,k}G(i\omega,k)
=\frac{1}{\Omega}\!\sum_k\!
\left [
\frac{\epsilon(k)}{2E_k}\coth\frac{E_k}{2 T}\!-\!\frac{1}{2}
\right ].
\label{n}
\ee
For $E_0=0$ the ground state will give the dominant contribution to the total density, i.e., there is a condensation of quasiparticles. A comparison with (\ref{dispersion}) and (\ref{qp}) yields the condition $\mu=\mu_0-\Sigma_{\rm reg}[\epsilon(0),0]=-\Delta_0(0)$ which corresponds to the Hugenholtz-Pines theorem \cite{HP59}, the condition for the usual Bose condensation.

So far we have only applied the Galitskii-Feynman and a quasi-particle approximation. To calculate $\omega_0$, $\mu$ and $\Delta_0(k)$ further approximations and the specification of the interaction are necessary. For the sake of simplicity we assume an s-type separable interaction
$V_q(p,k)=\lambda g_{p-\frac{q}{2}}g_{k-\frac{q}{2}}$
with Yamaguchi form factors \cite{Y54}, 
$g_p=\gamma^2/(\gamma^2+{p^2})$, where the range of the potential is controlled by $\gamma$ while $\lambda$ represents the strength of the interaction. We also neglect the higher orders in $k$ on the right hand side of (\ref{qp}). For a separable interaction the T-matrix can be expressed as
\be
{\cal T}_{q}(iz,p,k)=\frac{\lambda g_{p-\frac{q}{2}}g_{k-\frac{q}{2}}}{1+\frac{\lambda}{\Omega}\sum\limits_{k'}g_{k'-\frac{q}{2}}^2{\cal G}_q(iz,k')}.
\label{t}
\ee
From Eq.~(\ref{dq2}) follows $\Delta_0(k)=\Delta\, g_k$. The T-matrix (\ref{t}) diverges giving a bound state only if $\lambda<0$, therefore the condensation of bound states as mentioned above is possible only for attractive interaction. 
Now we are ready to perform the limit of infinite volume $\Omega$. The momentum sums will turn into integrals $\sum_k\to\Omega\int d^3k/(2\pi)^3$, and the singular contributions have to be considered separately. The condensate density $n_0$ and quantities $\omega_0$, $\mu$, $\Delta$ can be determined for a given temperature, total density and interaction strength with the help of two subsidiary functions:
\be
n_{\rm gas}(\mu,\Delta)&=&\int\frac{d^3k}{(2\pi)^3}\left[\frac{\epsilon(k)}{2E_k}\coth\frac{ E_k}{2T}-\frac{1}{2}\right],
\label{15}
\\
\lambda_b(\omega_0,\mu,\Delta)&=&\!\!-\left[\int\!\! \frac{d^3k}{(2\pi)^3}g_{k}^2{\cal G}_0(\omega_0\!-\!2\mu,k)\right]^{-1}\!.
\label{16}
\ee
Here $n_{\rm gas}$ is the density of the non-condensed Bose gas and $\lambda_b$ is the interaction strength which would create a bound state of energy $\omega_0$. We will use these functions to discuss the phase diagram.

\begin{figure}[]
\psfig{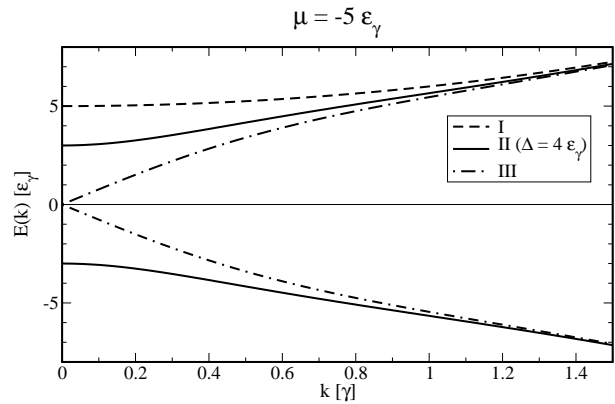} 
\caption{Single-particle dispersion for the three different cases: (I) quasi ideal, (II) gap, (III) gap-less and linear dispersion. The energy scale is given by the interaction range, $\varepsilon_\gamma=\hbar^2\gamma^2/2m$.}
\label{disp}
\end{figure}

\begin{figure}[]
\psfig{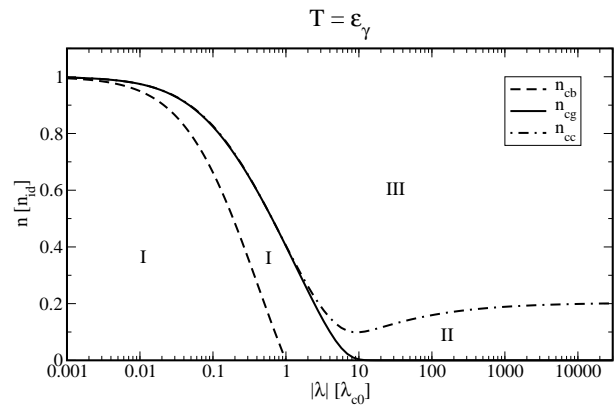} 
\caption{Phase diagram for the three different cases. The critical densities are given in units of the ideal critical density for Bose condensation $n_{\rm id}$ and the couplings strength in units of the critical interaction strength in vacuum $\lambda_{c0}$. Above $n_{\rm cb}$ the formation of bound pairs is possible. At $n_{\rm cg}$ appears the transition from the quasi ideal phase (I) to the Evans-Rashid phase (II) with condensate of bound pairs. Above $n_{\rm cc}$ there is an additional condensation of quasi particles, i.e., the usual Bose condensation phase (III).}
\label{phased}
\end{figure}

\begin{figure}[]
\psfig{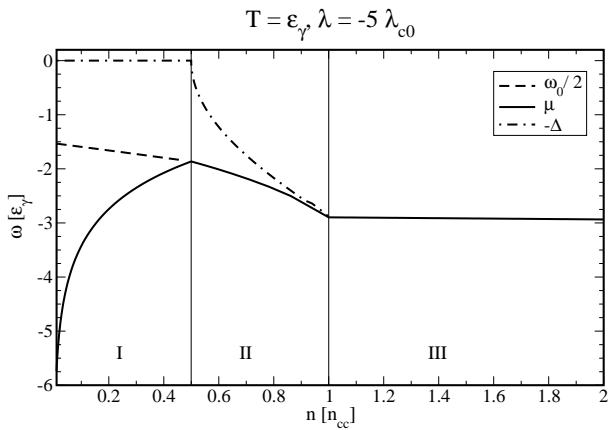} 
\caption{The binding energy, chemical potential and $\Delta$ in the three different regions according to a cut through the phase diagram in Fig.~\ref{phased} at $\lambda=-5 \lambda_{c0}$.}
\label{ebmd}
\end{figure}

As already mentioned there are three phases (I-III). The single-particle dispersions (\ref{dispersion}) at phases (I-III) are shown in Fig.~\ref{disp}. Critical lines between the low, medium and high density phases (I), (II) and (III) are specified in the phase diagram of Fig.~\ref{phased}. Figure~\ref{ebmd} shows the density dependence of $\omega_0$, $\mu$ and $\Delta$ for $\lambda=-5\lambda_{c0}$. In the following we discuss these results in more detail.
According to Eq.~(\ref{dq2}) the $\Delta$ vanishes as $1/\Omega$ for $\omega_0>2\mu$. The dispersion (\ref{dispersion}) is thus quadratic leading to the ideal gas behavior. Since $E_0=-\mu>0$ there is no Bose condensate, $n_0=0$. The chemical potential is determined from the total density $n$ via $n=n_{\rm gas}(\mu,0)$. The binding energy follows from $\lambda=\lambda_b(\omega_0,\mu,0)$. The two-particle propagator in this case is
${\cal G}_0(i\omega,k)={\coth[{\epsilon(k)}/{2T}]}/{[2\epsilon(k)-i\omega]}$. In vacuum the bound state vanishes for ${\lambda>-\lambda_{c0}=-\frac{8\pi\hbar^2}{m\gamma}}$. For $|\lambda|<\lambda_{c0}$ the bound state appears at the critical density $n_{cb}$ given by $\lambda=\lambda_b[0,\mu(n_{cb}),0]$. The critical density $n_{cb}$ is the dashed line in the phase diagram Fig.~\ref{phased}. As already mentioned $\omega_0$ and $2\mu$ move towards each other with increasing $n$. At $\omega_0=2\mu$ the condensation of bound states starts driving the Evans-Rashid transition \cite{ER73,S94}. The critical density of the transition $n_{cg}$ follows from $\lambda=\lambda_b[2\mu(n_{cg}),\mu(n_{cg}),0]$. In the phase diagram of Fig.~\ref{phased} the critical density $n_{cg}$ is denoted by the solid line. 
Above $n_{cg}$ a finite $\Delta$ appears due to the condensation of bound pairs. The chemical potential $\mu$ is then pinned to the pair condensate $\omega_0=2\mu$. The dispersion now consists of two branches separated by a gap $2\sqrt{\mu^2-\Delta^2}$. The gap implies $n_0=0$. The values of $\mu$ and $\Delta$ follow from the equations $n=n_{\rm gas}(\mu,\Delta)$ and $\lambda=\lambda_b(2\mu,\mu,\Delta)$. In this case only the two-particle propagator (\ref{g2pns}),
${{\cal G}_0(0,p)={\coth({E_p}/{2T})}/{2 E_p}}$, is needed. 
With increasing density, $\Delta$ increases and $\mu$ decreases. As can be seen in Fig.~\ref{ebmd}, their absolute values approach each other, i.e., the gap decreases. At $\mu=-\Delta$ the gap vanishes and the condensation of quasiparticles starts. The critical density of this transition is $n_{cc}=n_{\rm gas}(-\Delta,\Delta)$ with $\Delta$ given by $\lambda=\lambda_b(-2\Delta,-\Delta,\Delta)$. 
Due to the condensation of quasiparticles the zero momentum contribution to the density (\ref{n}) has to be split off and has to be considered separately as condensate density $n_0$. The chemical potential is pinned to the condensate, $\mu=-\Delta$ and $\omega_0=-2\Delta$. In analogy to the density also the zero momentum component of the sum in (\ref{t}) has to be split off. Therefore $n_0$ and $\Delta$ follow from the equations $n=n_0+n_{\rm gas}(-\Delta,\Delta)$ and ${1/\lambda=-n_0/\Delta+1/\lambda_b(-2\Delta,-\Delta,\Delta)}$. The dispersion (\ref{dispersion}) becomes gap-less and linear for small momenta,
\be
E_k=\sqrt{\frac{\Delta}{m}+\frac{2\Delta^2}{\hbar^2\gamma^2}}\hbar k+{\cal O}(k^2).
\ee
In leading order of the interaction, i.e., for small $\lambda$ and $\Delta$, we have $\Delta=-\lambda n_0$. In the limit of contact interaction, $\gamma\to\infty$, one then recovers the well known Bogoliubov dispersion
\be
E_k=\sqrt{\left(\frac{\hbar^2k^2}{2m}\right)^2\!\!-\!\lambda n_0\frac{\hbar^2k^2}{m}}=\sqrt{\!\frac{-\lambda n_0}{m}}\hbar k\!+\!{\cal O}(k^2)
\ee
leading to a positive and real sound velocity.

\begin{figure}[]
\psfig{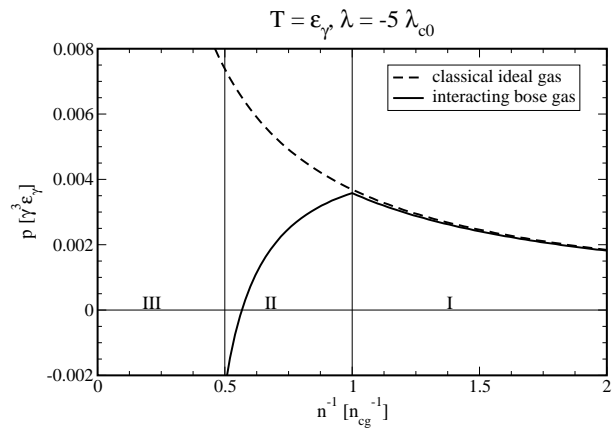} 
\caption{The pressure versus inverse density in the three different phases compared to the ideal gas behavior, $p=nT$.
}
\label{p}
\end{figure}

Finally we have to check whether the system is stable or not. To this end  we present in Fig.~\ref{p} the pressure evaluated via
\be
p_2-p_1=\int\limits_{\mu_1}^{\mu_2}n[\mu',\Delta(\mu')]d\mu'.
\ee 
The pressure shows a kink at the Evans-Rashid transition which is therefore of second order as well as the transition to Bose condensation of atoms. Due to the attractive interaction the pressure decreases above the Evans-Rashid transition density $n_{cg}$. The interaction potential does not include short range repulsion or a hard core which leads to the artifact that the decrease of the pressure cannot be compensated. As the density increases further the resulting pressure becomes negative leading to the collapse of the system. Assuming a Van-der-Waals behavior of the pressure one would expect the condensation into a liquid to start already below $n_{cg}$. The exact description of this condensation line is still open, however. Naturally, the theory of three phases described above is plausible only for positive pressure. On the other hand the usual densities in the experiment are so low, that the condensation into a liquid can be avoided during the experiment.

\begin{table}[t]
\caption{Interaction parameters and the typical energy and temperature scale for $^1$H and $^4$He}
\label{params}
\begin{tabular}{ccccccc}
\hline
\hline
&&&&&&\\[-1ex]
&&$\gamma$&&$\lambda/\lambda_{c0}$&&$\epsilon_\gamma/k_B$\\
&&(\AA$^{-1}$)&&&&(K)\\
\hline
&&&&&&\\[-1ex]
$^{1}$H&&$0.7031$&&$-2261$&&$12$\\
$^{4}$He&&$0.3962$&&$-1.057$&&$0.95$\\
\hline
\hline
\end{tabular}
\end{table}

We restricted ourselves in this paper to the case of bosons with zero total spin, e.g. $^1$H and $^4$He. To obtain the interaction parameters $\lambda$ and $\gamma$ as shown in Table~\ref{params} a comparison with numerical or experimental data is necessary. For $^4$He $\lambda$ and $\gamma$ were derived form the scattering length $a_0=93${\AA} \cite{JA95} and the effective range $r_0=7.298${\AA} \cite{JA95}. In this way one can also obtain the vacuum binding-energy $\omega_{00}=129.6$neV and bond length $\langle r\rangle=49${\AA}, which are in good agreement with the numerical and experimental data $\omega_{00}=(130\pm7)$neV \cite{JA95} and $\langle r\rangle=(52\pm4)${\AA} \cite{GSTHKS00}, respectively. For $^1$H the dissociation energy $D_0=4.478$eV \cite{CRC08} and the bond length $\langle r\rangle=0.7414${\AA} \cite{CRC08} were used to fit the potential describing the $^1$H$_2$ ground state.

Our results are very similar to those obtained by Stoof. We find two phase transitions. The first one is the Evans-Rashid transition between a quasi-ideal gas and a BCS-like phase connected with a condensation of bound states. The second phase transitions is the onset of the usual Bose condensation of quasiparticles. We obtain the expected linear dispersion in the Bose condensate region without the use of anomalous propagators. The kinks in the chemical potential and the pressure indicate that the Evans-Rashid transition as well as the transition to the phase of condensed atoms are of second order. The decrease of the pressure above $n_{cg}$ and the change of its sign are a strong signatures of a gas-liquid or gas-solid transition and therefore the quantum phase transitions might be inaccessible. Concluding we have employed a novel many-body correction scheme for the T-matrix which allows us to describe multiple condensed phases as well as gaped phases simultaneously on the same theoretical foundation.

This work was supported by DFG 
Priority Program 1157 via GE1202/06, the BMBF and by European 
ESF program NES as well as
Czech research plan MSM 0021620834. The financial support by the Brazilian Ministry of Science and Technology is acknowledged.

\bibliography{kmsr,kmsr1,kmsr2,kmsr3,kmsr4,kmsr5,kmsr6,kmsr7,delay2,spin,gdr,refer,sem1,sem2,sem3,micha,genn,solid,deform,bose,delay3,paradox}

\end{document}